\documentclass[twocolumn,prl,aps]{revtex4}

\newcommand{\be}{\begin{equation}} \newcommand{\ee}{\end{equation}} 
\newcommand{\bea}{\begin{eqnarray}}\newcommand{\eea}{\end{eqnarray}}

\begin{document}
\preprint{ 
%VB/Physics/09-xx, 
quant-ph/0901.1730}
\title{ Quantum Phase Transition in a Pseudo-hermitian Dicke model}
\author{ Tetsuo Deguchi$^1$} \email{deguchi@phys.ocha.ac.jp}
\author{ Pijush K. Ghosh$^2$} \email{pijushkanti.ghosh@visva-bharati.ac.in}
\affiliation{ $^1$ Department of Physics, Graduate School of Humanities and
Sciences,\\ Ochanomizu University,
2-1-1 Ohtsuka, Bunkyo-ku,\\ Tokyo 112-8610, Japan\\
%\author{ Pijush K. Ghosh} \email{pijushkanti.ghosh@visva-bharati.ac.in}
%\affiliation{
$^2$ Department of Physics, Siksha-Bhavana,\\ 
Visva-Bharati University,\\
Santiniketan, PIN 731 235, India.}
\begin{abstract} 
We show that a Dicke-type non-hermitian Hamiltonian admits entirely
real spectra by mapping it to the ``Dressed Dicke Model"(DDM)
through a similarity transformation. We find a positive-definite
metric in the Hilbert space of the non-hermitian Hamiltonian so
that the time-evolution is unitary and allows a consistent quantum
description. We then show that this non-hermitian Hamiltonian describing
non-dissipative quantum processes undergoes Quantum Phase Transition(QPT).
The exactly solvable limit of the non-hermitian Hamiltonian has also
been discussed.
\end{abstract}
\pacs{PACS numbers: 03.65.-w, 03.65.Fd, 73.43.Nq, 42.50.Ct }
\maketitle

Although the choice of a proper set of hermitian operators is sufficient
to ensure the reality of the entire spectra and unitary time-evolution 
for a quantum system, it is neither necessary nor dictated by any
fundamental principle. 
It is known since the pioneering work of Bender and Boettcher\cite{bend}
that ${\cal{PT}}$-symmetric non-hermitian operators
with an appropriate inner-product in the Hilbert space give consistent
description of non-dissipative quantum processes.
The Hamiltonian that is non-hermitian with respect to the conventional
inner-product in the Hilbert space becomes hermitian with respect to
the new inner product and results of a hermitian theory follow naturally.
The same problem can be studied using pseudo-hermitian
operator\cite{ali,quasi}, i.e. an operator that is related to its adjoint
through a similarity transformation. Both the approaches involving
pseudo-hermiticity and ${\cal{PT}}$-invariance are complementary to
each other and open up several new directions in the study of
non-hermitian operators
\cite{bend,ali,quasi,ddt,khare,wadati,bpm,bm,kumar,nc,swan,me,quesne,smilga}.
It may be mentioned here that operators which are non-hermitian with respect
to the conventional inner product in the Hilbert space are generally used
to simulate dissipative processes. In this letter, we are concerned
about a subclass of such non-hermitian operators which are also pseudo-hermitian
and may be used consistently to describe non-dissipative processes with a
modified inner product in the Hilbert space.

The study on QPT\cite{ss} has received considerable attention in recent
times and reveals many aspects that are qualitatively different from that
of phase transition at finite temperature. The investigations so far are mainly
restricted to hermitian Hamiltonian,
since an entirely real spectra with a well-defined ground-state is not
guaranteed a priory for a non-hermitian Hamiltonian. The dynamics of QPT in
a closed system is governed by non-dissipative terms,
since the system at zero temperature is already in thermal equilibrium.
Unlike the phase-transitions at finite temperature, the time-evolution from
one phase to the other is expected to be unitary for a system undergoing QPT.
It is thus nontrivial for a non-hermitian Hamiltonian to describe a QPT in a
closed system, since the time-evolution is not necessarily unitary.
It is natural to ask at this juncture whether or not one could discuss
about QPT in a closed system within the framework of ${\cal{PT}}$-symmetric
non-hermitian Hamiltonian. If the answer is in the affirmative, 
it may open up new directions in the study of several interlinked
areas of physics like level statistics, quantum entanglement, quantum chaos
etc. within the framework of pseudo-hermitian and/or ${\cal{PT}}$-symmetric
non-hermitian Hamiltonian. The enlarged
parameter-space of a non-hermitian Hamiltonian compared to its hermitian
counterpart may prove to be an added advantage.

One of the main results of this letter is that a pseudo-hermitian
deformation of the DDM indeed undergoes QPT. We consider the 
non-hermitian Dicke-type Hamiltonian\cite{dicke},
\bea
H & = & \omega a^{\dagger} a + \theta_1 e^{i \xi_1} a^2 +
\theta_2 e^{-i \xi_1} {a^{\dagger}}^2 
+ \alpha e^{i \xi_2} J_-  a^{\dagger}\nonumber \\
& + & \beta  e^{- i \xi_2} J_+ a
+ \gamma e^{i \xi_3} J_- a + \delta e^{-i \xi_3} J_+ a^{\dagger}
+ \omega_0 J_z,
\label{eq1}
\eea
\noindent where $\omega, \omega_0, \theta_1$, $\theta_2$, 
$\alpha, \beta, \gamma, \delta, \xi_1, \xi_2, \xi_3$ are real parameters;
$a$, $a^{\dagger}$ are the standard bosonic annihilation-creation
operators and $J_{z}$, $J_{\pm}:= J_x \pm i J_y$ are the generators of
the $SU(2)$ algebra,
\bea
&& \left [ a, a^{\dagger} \right ] =1, \nonumber \\
&& \left [ J_+, J_- \right ] = 2 J_z, \ \ \ \ \left [ J_z, J_{\pm} \right ]
\ = \pm J_{\pm}.
\eea
\noindent The Hamiltonian $H$ commutes with the parity operator $\Pi$,
\be
\Pi = e^{i \pi \hat{N}}, \ \ \hat{N} = a^{\dagger} a + J_z + j,
\ee
\noindent where $j$ is the total spin-angular momentum. The eigenstates of
$H$ have definite parity depending on whether the eigenvalues of the
operator $\hat{N}$ are odd or even.
In general, the Hamiltonian $H$ is non-hermitian. The Hermitian
Hamiltonian is obtained in the limit,
\be
\alpha = \beta, \gamma = \delta, \theta_1 = \theta_2,
\ee
\noindent and is known as the DDM in the literature\cite{djc,sjc}.
The standard Dicke model is obtained by a further choice of
$\theta_1=\theta_2=\xi_1=\xi_2=\xi_3=0$ and $\alpha=\beta=\gamma=\delta$.
The Dicke Hamiltonian has been studied extensively from the viewpoint of
QPT\cite{lieb,hillery,emary}, level-statistics\cite{emary}, quantum
entanglement\cite{lambert,buzek} and exact solvability\cite{hikami}.
Certain spintronics based models\cite{hikami,datta} with Dresselhaus
and Rashba-type spin-orbit interactions can be mapped to the 
Dicke model, implying its relevance in the study of two dimensional
semi-conductor physics. The Tavis-Cummings model\cite{tc} is obtained in the
limit $\theta_1=\theta_2=\gamma= \delta=0$ and it reduces to
the Jaynes-Cummings model\cite{jc} if the fundamental
representation of the $SU(2)$ is used. Non-hermitian versions of both
the Tavis-Cummings and the Jaynes-Cummings models have been studied
previously\cite{me}.
The Hamiltonian with $\omega_0=\alpha=\beta=\gamma=\delta=0$ is known as
the `Swanson model'\cite{swan} in the context of ${\cal{PT}}$-symmetric
quantum mechanics and has been studied in some detail\cite{swan,quesne}.
In this letter, we study the Hamiltonian $H$ with its full generality
and show the existence of QPT for certain special choices of the parameters.

The Hamiltonian $H$ can be mapped to a hermitian Hamiltonian ${\cal{H}}$
through a similarity transformation when the following relations are satisfied,
\bea
&& \alpha \ \delta - \beta \ \gamma =0, \ \ \theta_1= \theta_2=0 \nonumber \\
&& \alpha \ \delta \ \theta_1 - \beta \ \gamma \ \theta_2 = 0,
\ \ \theta_1 \neq 0 \neq \theta_2.
\label{pseudocon}
\eea
\noindent To see this, define an operator $\rho$ and its inverse as,
\bea
&& \rho = e^{\hat{O}}, \ \ \ \rho^{-1} = e^{-\hat{O}},\nonumber \\
&& \hat{O} =  \frac{1}{4} ln \left ( \frac{\theta_1}{\theta_2} \right )
a^{\dagger} a
+ \frac{1}{4} ln \left ( \frac{\alpha \gamma}{\beta \delta} \right ) 
\left ( J_z + j \right ).
\eea
\noindent The operator $\rho$ is positive-definite and well-defined provided
the following relations are satisfied,
\be
\frac{\theta_1}{\theta_2} > 0, \ \
\frac{\alpha}{\beta} > 0, \ \
\frac{\gamma}{\delta} > 0. \ \
\label{eqp}
\ee 
\noindent The conditions $\frac{\theta_1}{\theta_2} > 0$ and
$\frac{\alpha \gamma}{\beta \delta} > 0$ are sufficient to ensure that
$\rho$ has the desired property. The much more stringent
condition(\ref{eqp}) is used to make the transformed Hamiltonian
${\cal{H}}$ hermitian.
The operator $\hat{O}$ can be constructed for several special cases
as follows:
\bea
&& \hat{O} =  \frac{1}{4} ln \left ( \frac{\theta_1}{\theta_2} \right )
a^{\dagger} a, \ \
\frac{\theta_1}{\theta_2} > 0, \ \ \alpha= \beta=\gamma=\delta=0;\nonumber \\
&& \hat{O} = \frac{1}{4} ln \left ( \frac{\alpha \gamma}{\beta \delta}
\right ) \left ( J_z + j \right ),\nonumber \\
&& \theta_1=\theta_2=0, \ \ \frac{\alpha}{\beta} > 0, \ \
\frac{\gamma}{\delta} > 0;\nonumber \\
&& \hat{O} =  \frac{1}{4} ln \left ( \frac{\theta_1}{\theta_2} \right )
a^{\dagger} a + \frac{1}{4} ln \left ( \frac{\alpha}{\beta} \right )
\left ( J_z + j \right ),\nonumber \\
&& \frac{\theta_1}{\theta_2} > 0, \ \ \frac{\alpha}{\beta} > 0, \ \ \gamma=
\delta=0;\nonumber \\
&& \hat{O} =  \frac{1}{4} ln \left ( \frac{\theta_1}{\theta_2} \right )
a^{\dagger} a
+ \frac{1}{4} ln \left ( \frac{\gamma}{\delta} \right ) 
\left ( J_z + j \right ),\nonumber \\
&& \frac{\theta_1}{\theta_2} > 0, \ \ \frac{\gamma}{\delta} > 0,  \ \ \alpha=
\beta=0.
\eea
\noindent We will be working within the range of the parameters defined
by Eq. (\ref{eqp}) unless mentioned otherwise.
Using the Baker-Campbell-Hausdorff formula,
\be
e^ A B e^{-A} = B + [A, B] + \frac{1}{2!} [A, [A, B]] 
+ \frac{1}{3!} [A, [A, [A, B]]] + \dots,
\ee
we find,
\bea
{\cal{H}} & = & \rho H \rho^{-1}\nonumber \\
&=& \omega a^{\dagger} a + \sqrt{\theta_1 \theta_2}
\left ( e^{i \xi_1} \ a^2 + e^{- i \xi_1} \ {a^{\dagger}}^2 \right )
+ \omega_0 J_z\nonumber \\
& + & \sqrt{ \alpha \beta} \left ( e^{i \xi_2} \ J_-  a^{\dagger}
+ e^{-i \xi_2} \ J_+ a \right )\nonumber \\
& + & \sqrt{\gamma \delta}
\left ( e^{i \xi_3} \ J_- a +  e^{-i \xi_3} \ J_+ a^{\dagger} \right ),
\label{original}
\eea
\noindent when the condition (\ref{pseudocon}) is satisfied.
Note that ${\cal{H}}$ is hermitian, since $\theta_1 \theta_2$,
$\alpha \beta$ and $\gamma \delta$ are positive-definite due to the
condition (\ref{eqp}). The Hamiltonian $H$ is quasi-hermitian, i.e.,
related to the hermitian Hamiltonian ${\cal{H}}$ through a similarity
transformation. The pseudo-hermiticity of $H$,
i.e. $H^{\dagger} = \eta_+ H \eta_+^{-1}$, follows automatically
where the metric $\eta_+$ is given by $\eta_+:= \rho^2$.
The Hamiltonian $H$ that is non-hermitian under the Dirac-hermiticity
condition becomes hermitian with respect to the modified inner-product
defined in the Hilbert space as,
$\langle \langle u,v\rangle \rangle_{\eta_+} := \langle u,\eta_+ v \rangle$.
In particular,
\be
\langle u | H v \rangle \neq \langle H u | v \rangle, \ \
\langle \langle u|H v \rangle \rangle_{\eta_+} = \langle \langle H u | v
\rangle \rangle_{\eta_+}.
\ee
\noindent Thus, with the modified inner-product, the results of a hermitian
Hamiltonian follow automatically.

A comment is in order at this point. The atomic inversion and the mean
photon number are determined by the expectation values of the operators
$J_z$ and $a^{\dagger} a$, respectively. Both the operators $J_z$ and
$a^{\dagger} a$ commute with $\eta_+$ and hence, are hermitian with respect
to the modified inner-product. However, operators like $J_x$, $J_y$,
$a + a^{\dagger}$ and $i (a^{\dagger} - a)$, which are hermitian with respect
to the Dirac-hermiticity condition, are no longer hermitian with respect
to the modified inner product. It may be noted here that corresponding to each
operator ${\cal{A}}$ that is hermitian with respect to the Dirac-hermiticity
condition, the operator $\hat{\cal{A}} := \rho^{-1} {\cal{A}} \rho$ is
hermitian with respect to the modified inner product\cite{ali}. Consequently,
the operator $\hat{\cal{A}}$ is a physical observable in the Hilbert space
of $H$ that is endowed with the metric $\eta_+$. Following this prescription,
a set of $SU(2)$ generators those are hermitian with respect to the modified
inner-product can be constructed as follows:
\bea
\hat{J}_x & := & J_x cosh \Gamma - i J_y sinh \Gamma\nonumber \\
\hat{J}_y & := & J_y cosh \Gamma + i J_x sinh \Gamma\nonumber \\
\hat{J}_z & := & J_z, \ \ \ \
\Gamma \equiv \frac{1}{4} ln \left ( \frac{\alpha \gamma}{\beta \delta}
\right ). 
\eea
\noindent Similarly, annihilation operator $\hat{a}$ and its adjoint
$\hat{a}^{\dagger}$ can be obtained as,
\be
\hat{a} := \left ( \frac{\theta_1}{\theta_2} \right )^{\frac{1}{4}} a, \ \ \ \
\hat{a}^{\dagger} := \left ( \frac{\theta_1}{\theta_2}
\right )^{-\frac{1}{4}} a^{\dagger}. \ \ \ \
\ee
\noindent The non-hermitian Hamiltonian $H$ can be re-written in terms of
these operators as,
\bea
H &=& \omega \hat{a}^{\dagger} \hat{a} + \sqrt{\theta_1 \theta_2}
\left ( e^{i \xi_1} \ \hat{a}^2 + e^{- i \xi_1} \ (\hat{a}^{\dagger})^2 \right )
+ \omega_0 \hat{J}_z\nonumber \\
& + & \sqrt{ \alpha \beta} \left ( e^{i \xi_2} \ \hat{J}_-  \hat{a}^{\dagger}
+ e^{-i \xi_2} \ \hat{J}_+ \hat{a} \right )\nonumber \\
& + & \sqrt{\gamma \delta}
\left ( e^{i \xi_3} \ \hat{J}_- \hat{a} +  e^{-i \xi_3} \
\hat{J}_+ \hat{a}^{\dagger} \right ),
\label{original}
\eea
\noindent where $\hat{J}_{\pm} := \hat{J}_x \pm i \hat{J}_y$.

The hermitian Hamiltonian ${\cal{H}}$ has the form of the DDM
and has been extensively studied in the literature\cite{djc,sjc}.
In general, the Hamiltonian ${\cal{H}}$ is not exactly solvable.
Using the Bogoliubov transformation,
\be
\left ( \matrix{ {b}\cr \\ {{b}^{\dagger}}} \right ) =
\left ( \matrix{{cos h \theta} & {e^{i \phi} sin h \theta}\cr \\
{e^{- i \phi} sin h \theta} & {cos h \theta}} \right )
\left ( \matrix{{a}\cr \\ {a^{\dagger}}} \right ), 
\ee
\noindent either the counter-rotating terms $J_ - a, J_+ a^{\dagger}$
or the double frequency terms $a^2, {a^{\dagger}}^2$ in the Hamiltonian
${\cal{H}}$ can be eliminated with all other terms appearing with
renormalized coupling constants. Both the counter-rotating and the
double frequency terms can be eliminated simultaneously for
fixed values of $\theta$ and $\phi$, if a constraint involving the parameters
$\alpha, \beta, \gamma, \delta, \theta_1$ and $\theta_2$ is also satisfied.
Let us choose $\phi$ and $\theta$ as,
\bea
&& \phi= - \xi_1, \ \ \theta= tanh^{-1} \left (
\frac{\Delta}{2 \sqrt{\theta_1 \theta_2}} \right );\nonumber \\
&& \Delta \equiv \omega - \left ( \omega^2 - 4 \theta_1 \theta_2 \right
)^{\frac{1}{2}}, \ \ \ \ \omega^2 > 4 \theta_1 \theta_2,
\eea
\noindent so that the double-frequency terms are eliminated from ${\cal{H}}$.
It may be mentioned here that the choice of $\phi$ and $\theta$ as,
\bea
&& \phi= - \xi_1, \ \ \tilde{\theta}= tanh^{-1} \left (
\frac{\tilde{\Delta}} {2 \sqrt{\theta_1 \theta_2}} \right );\nonumber \\
&& \tilde{\Delta} \equiv \omega + \left ( \omega^2 - 4 \theta_1 \theta_2 \right
)^{\frac{1}{2}}, \ \ \ \ \omega^2 > 4 \theta_1 \theta_2,
\eea
\noindent also removes the double-frequency terms. However, this solution
leads to unphysical situations and is discarded henceforth. 
Using the condition (\ref{pseudocon}) and demanding the removal of
counter-rotating terms,
the values of $\gamma$, $\delta$ and $\xi_3$ are determined as, 
\be
\gamma = \frac{\alpha \Delta}{2 \theta_2}, \ \ \ \ 
\delta = \frac{\beta \Delta}{2 \theta_1},\ \ \ \
\xi_3 = \xi_1 + \xi_2.
\ee
\noindent The Hamiltonian ${\cal{H}}$
can be expressed in terms of the new canonical operators $b, b^{\dagger}$
as,
\bea
&& {\cal{H}} = \omega_0 J_z + \Omega b^{\dagger} b + \Omega_0
+ \Omega_1 \left ( e^{-i \xi_2} \ J_+ b + e^{i \xi_2} \
J_- b^{\dagger} \right ),\nonumber \\
&& \Omega_0 = -\frac{\Delta}{2}, \ \ \ 
\Omega = \frac{\left(\omega^2 -4 \theta_1 \theta_2 \right) \Delta }{
4 \theta_1 \theta_2 - \omega \Delta},\nonumber \\
&& \Omega_1= \sqrt{\frac{\alpha \beta}{2 \theta_1 \theta_2}}
\left ( 4 \theta_1 \theta_2 - \omega \Delta \right )^{\frac{1}{2}},
\label{tc}
\eea
\noindent which has the form of the Tavis-Cummings model or the
DDM in the rotating-wave approximation with the
modified coupling constants. All these coupling constants
$\Omega_0$, $\Omega$ and $\Omega_1$ are real and $\Omega$ is also
positive-definite for $\omega^2 > 4 \theta_1 \theta_2$.

The Hamiltonian ${\cal{H}}$ in Eq. (\ref{tc}) is exactly
solvable\cite{hillery}. It can be decomposed in terms of two
mutually commuting operators $K$ and $L$ as follows,
\bea
&& {\cal{H}} = \Omega K + \Omega_1 L + \Omega_0,\nonumber \\
&& K = b^{\dagger} b + J_z,\nonumber \\
&& L = e^{-i \xi_2} \ J_+ b +
e^{i \xi_2} \ J_- b^{\dagger}
+ \frac{\omega_0 - \Omega}{\Omega_1} J_z.
\eea
\noindent The operator $K$ is diagonal for a fixed spin $j$
with the eigenvalues of $J_z$ as $(m-j)$, $m=0, 1, \dots 2j$ and that
of the bosonic number operator $b^{\dagger} b$ as $n$, $n=0, 1, 2 \dots$.
The operator $L$ and hence, the operator ${\cal{H}}$
can be diagonalized in the basis spanned by the eigenstates of $K$.
Let $|n,m;j\rangle_{\cal{H}}$ be a complete set of orthonormal eigenstates
of ${\cal{H}}$ with the eigenvalues $E_{n,m;j}$. The orthonormality
of $|n,m;j\rangle_{\cal{H}}$ is based on the standard inner-product in
the Hilbert space. The eigenstates of $H$ with the same eigenvalues
$E_{n,m;j}$ are determined as,
\be
|n,m;j\rangle_{H} = \rho^{-1} \ \ |n,m;j\rangle_{\cal{H}},
\ee
\noindent which form a complete set of orthonormal eigenstates 
under the modified inner-product defined in the Hilbert space of $H$.
Consequently, the non-hermitian Hamiltonian $H$ is also exactly solvable
and admits consistent quantum description.

The expectation value of an
operator $X$ in the Hilbert space of $H$ is determined as,
\be
\langle \langle X \rangle \rangle_{\eta_+} = \langle n, m;j | \rho
X \rho^{-1} | n, m; j \rangle_{\cal{H}}.
\ee
\noindent Both $J_z$ and $a^{\dagger} a$ are hermitian with respect
to the Dirac-hermiticity condition as well as with respect to
the modified inner product. In particular, both $J_z$ and $a^{\dagger} a$
commute with $\rho$, leading to the results:
\bea
&& \langle \langle J_z \rangle \rangle_{\eta_+} =
\langle n, m;j | J_z | n, m; j \rangle_{\cal{H}},\nonumber \\
&& \langle \langle a^{\dagger} a \rangle \rangle_{\eta_+} =
\langle n, m;j | a^{\dagger} a | n, m; j \rangle_{\cal{H}}.
\eea
\noindent  Thus, both
$\langle \langle J_z \rangle \rangle_{\eta_+}$ and
$\langle \langle a^{\dagger} a \rangle \rangle_{\eta_+}$ are real.
However, in general, $\langle \langle J_x \rangle \rangle_{\eta_+}$, 
$\langle \langle J_y \rangle \rangle_{\eta_+}$, 
$\langle \langle a + a^{\dagger} \rangle \rangle_{\eta_+}$ and 
$\langle \langle i (a^{\dagger} - a ) \rangle \rangle_{\eta_+}$
are complex,
\bea
&& \langle \langle J_x \rangle \rangle_{\eta_+} =
cosh \Gamma \langle J_x \rangle_{\cal{H}}
+ i sinh \Gamma \langle J_y \rangle_{\cal{H}} \nonumber \\
&& \langle \langle J_y \rangle \rangle_{\eta_+} =
- i sinh \Gamma \langle J_x \rangle_{\cal{H}}
+ cosh \Gamma \langle J_y \rangle_{\cal{H}}\nonumber \\
&& \langle \langle a + a^{\dagger} \rangle \rangle_{\eta_+} = 
\left ( \frac{\theta_1}{\theta_2} \right )^{-\frac{1}{4}}
\langle a \rangle_{\cal{H}}
+ \left ( \frac{\theta_1}{\theta_2} \right )^{\frac{1}{4}}
\langle a^{\dagger} \rangle_{\cal{H}}\nonumber \\
&& \langle \langle a^{\dagger} - a \rangle \rangle_{\eta_+} = 
\left ( \frac{\theta_1}{\theta_2} \right )^{\frac{1}{4}}
\langle a^{\dagger} \rangle_{\cal{H}}
- \left ( \frac{\theta_1}{\theta_2} \right )^{- \frac{1}{4}}
\langle a \rangle_{\cal{H}},
\eea
\noindent where
$\langle J_x \rangle_{\cal{H}}$ and $\langle J_y \rangle_{\cal{H}}$ are real,
while $\langle a \rangle_{\cal{H}}$ and $\langle a^{\dagger} \rangle_{\cal{H}}$
are complex. As discussed before, corresponding to each operator ${\cal{A}}$
in the Hilbert space of ${\cal{H}}$, the physical observable in the Hilbert
space of $H$ that is endowed with the metric $\eta_+$ is
$\hat{\cal{A}}= \rho^{-1} {\cal{A}} \rho$.
The expectation values of these capped operators are real,
since $\langle \langle \hat{\cal{A}} \rangle \rangle_{\eta_+} =
\langle n, m;j | {\cal{A}} | n, m; j \rangle_{\cal{H}}$. Thus,
a complete and consistent description of the pseudo-hermitian $H$ is 
allowed with the proper identification of the physical observables,

The Hamiltonian ${\cal{H}}$ in Eq. (\ref{tc}) is known to exhibit
QPT\cite{hillery,emary}. Although ${\cal{H}}$ is exactly solvable,
it becomes tedious to calculate the eigenspectra for large $j$.
The Holstein-Primakoff representation of the $SU(2)$ generators
\be
J_- = \left ( 2 j - \zeta^{\dagger} \zeta \right )^{\frac{1}{2}} \zeta, \ \
J_+ = \zeta^{\dagger} \left ( 2 j -
\zeta^{\dagger} \zeta \right )^{\frac{1}{2}}, \ \
J_z = \zeta^{\dagger} \zeta - j,\nonumber \\
\ee
\noindent where $\zeta, \zeta^{\dagger}$ are the bosonic annihilation and
creation operators satisfying $\left [\zeta, \zeta^{\dagger} \right ]=1$,
can be used to study the thermodynamic limit $j \rightarrow \infty$.
It is important to note that among the $SU(2)$ generators, only the
combination $J_z + j$ appears in the expressions of the parity operator
$\Pi$ and the similarity operator $\rho$. Consequently, $\rho$ and $\Pi$
are well-defined in the thermodynamic limit $j \rightarrow \infty$.
Following the standard method described in\cite{hillery,emary},
the normal phase of ${\cal{H}}$ can be found to be described in the range
$ \lambda_1 < \lambda_1^c \equiv \sqrt{\Omega \omega_0}$, while the
`super-radiant phase' is described in the range $\lambda_1 >
\lambda_1^c$, where $\lambda_1 \equiv \sqrt{2 j} \Omega_1$.
The Hamiltonian $H$ and ${\cal{H}}$ have the same eigenspectra, since they
are related to each other through a similarity transformation. Moreover,
note that the operators $\hat{O}$, $\rho$ and $\rho^{-1}$ are well-defined
in the thermodynamic limit $j \rightarrow \infty$. Thus, the Hamiltonian
$H$ also undergoes QPT with the normal phase described in the range
$ \lambda_1 < \lambda_1^c$, while the `super-radiant phase'
is described in the range $\lambda_1 > \lambda_1^c$. The values of the
mean photon number and the atomic inversion above the critical value
$\lambda_1^c$ can be determined as follows:
\bea
j^{-1} \langle \langle a^{\dagger} a \rangle \rangle_{\eta_+}
= \frac{1}{2} \left ( 1 - \frac{\omega_0^2 \Omega^2}{\Omega_1^4} \right )
\frac{\Omega_1^2}{\Omega^2}\nonumber \\
j^{-1} \langle \langle J_z \rangle \rangle_{\eta_+} =
- \frac{\omega_0 \Omega}{\Omega_1^2}; \ \ \lambda_1 > \lambda_1^c.
\eea
\noindent This is one of the main results of this letter.

The hermitian Hamiltonian ${\cal{H}}$ in Eq. (\ref{original}) with
$\theta_1=\theta_2=\xi_1=\xi_2=\xi_3=0$ and
$\sqrt{\gamma \delta}=\sqrt{\alpha \beta} \equiv \frac{\lambda_2}{\sqrt{2 j}}$
reduces to the `standard Dicke model' which is known to undergo QPT for
$ \lambda_2 > \lambda_2^c \equiv \frac{\sqrt{\omega \omega_0}}{2}$
\cite{emary}. The non-hermitian Hamiltonian $H$ in (\ref{eq1})
with $\theta_1=\theta_2=0$, $\xi_1=\xi_2=\xi_3=0$,
$\gamma= \pm \alpha$, $\delta = \pm \beta$,
\be
\tilde{H} = \omega \ a^{\dagger} a +  
+ \omega_0 \ J_z + \alpha \ J_-  a^{\dagger} +
\beta  \ J_+ a
\pm \alpha  J_- a \pm \beta J_+ a^{\dagger},
\ee
\noindent is equivalent to the `standard Dicke Model' through the similarity
transformation $H_{Dicke}= \rho \tilde{H} \rho^{-1}$ with the operator
$\hat{O}$ given by,
\be
\hat{O}=\frac{1}{2} ln ( \frac{\alpha}{\beta} ) \left (
J_z + j \right ), \ \ \frac{\alpha}{\beta} > 0.
\ee
\noindent Thus, the non-hermitian Hamiltonian $\tilde{H}$ also undergoes QPT
for $\lambda_2 > \lambda_2^c$. The values of the atomic inversion and the
mean photon number above the critical value $\lambda_2^c$ are identical to
that of the standard Dicke model:
\bea
&& j^{-1} \langle \langle J_z \rangle \rangle_{\eta_+} =
- \left ( \frac{\lambda_2^c}{\lambda_2} \right )^2,\nonumber \\
&& j^{-1} \langle \langle a^{\dagger} a \rangle \rangle_{\eta_+} =
\frac{2 \lambda_2^2}{\omega^2} \left [ 1 -
\left (\frac{\lambda_2^c}{\lambda_2} \right )^4 \right ], \
\lambda_2 > \lambda_2^c.
\eea
\noindent The results for finite $j$, as quoted in Ref. \cite{emary}
for $H_{Dicke}$, are equally applicable for $\tilde{H}$, since
$\langle \langle J_z \rangle \rangle_{\eta_+} =
\langle J_z \rangle_{H_{Dicke}}$ and
$\langle \langle a^{\dagger} a \rangle \rangle_{\eta_+}=\langle
a^{\dagger} a \rangle_{H_{Dicke}}$.

The Hamiltonian ${\cal{H}}$ in Eq. (\ref{original}) with its full
generality also undergoes QPT for ${\mid \mu \mid} < 1$, 
\be
\mu \equiv \frac{ \omega_0 \left ( \omega + 2 \sqrt{\theta_1 \theta_2}
\right ) }{\left (\lambda_3 + \lambda_4 \right )^2}, \ \
\lambda_3 \equiv \sqrt{\frac{\alpha \beta}{2 j}}, \ \
\lambda_4 \equiv \sqrt{\frac{\gamma \delta}{2 j}}.
\label{asli}
\ee
\noindent Consequently, $H$ with the parameters satisfying the relations
in Eq. (\ref{pseudocon}) also undergoes quantum phase transition for
${\mid \mu \mid} < 1$. The values of mean photon number and the atomic
inversion for $\mu < 1 $ can be determined as follows:
\bea
&& j^{-1} \langle \langle a^{\dagger} a \rangle \rangle_{\eta_+} =
\frac{1}{2} (1 - \mu^2) \left ( \frac{\lambda_3 + \lambda_4}{\omega +
2 \sqrt{\theta_1 \theta_2}} \right )^2 \nonumber \\
&& j^{-1} \langle \langle J_z \rangle \rangle_{\eta_+} =
 - \mu; \ \ \mu < 1.
 \label{1asli}
\eea
\noindent The mean photon number vanishes identically and the atomic inversion
is equal to $-1$ for $\mu > 1$. The QPT in the Tavis-Cummings model and the
Dicke model appear as special cases of the general result described
by Eqs. (\ref{asli}) and (\ref{1asli}).

We have shown that a non-hermitian version of the DDM undergoes QPT.
This is the first time in the literature that QPT for pseudo-hermitian
operators has been described and definitely broadens the scope of studying
QPT in various other non-hermitian models. For the particular case of
the pseudo-hermitian DDM, it is to be seen whether or not the QPT is related
to a change in level-statistics and/or cross-over from entangled to
disentangled states, as is the case for the standard Dicke
Hamiltonian\cite{emary,lambert}. Finally, as mentioned earlier, the DDM can be
mapped to certain spintronics-based models\cite{hikami,datta}. Our results
on QPT can be directly extended to such models and may prove to be the
testing ground of pseudo-hermitian quantum mechanics through
appropriate quantum engineering of two-dimensional semiconductor devices.

\acknowledgements{ We would like to thank K. Kudo for useful discussions.
PKG would like to thank Ochanomizu University for warm hospitality
during his visit under the JSPS Invitation Fellowship for Research in
Japan(S-08042), where a part of this work has been carried out.}

\end{document}